\documentclass[%
 reprint,
 amsmath,amssymb,
 aps,
]{revtex4-2}

\usepackage{graphicx}% Include figure files
\usepackage{dcolumn}% Align table columns on decimal point
\usepackage{bm}% bold math
\begin{document}

\preprint{APS/123-QED}

\title{Practical Derivations of Fermion and Gauge
Boson Reduction Formulae in Curved Spacetimes}% Force line breaks with \\

\author{Jesse Huhtala}
    \email{jejohuh@utu.fi}
\author{Iiro Vilja}%
 \email{vilja@utu.fi}
\affiliation{Finland, 20014 University of Turku, Faculty of Science,\\ Department of Physics and Astronomy, Laboratory of Theoretical Physics}

\date{\today}% It is always \today, today,

\begin{abstract}
LSZ-type reduction formulae are derived for gauge fields and fermions in curved spacetime. The formulae are derived using a conserved current method applicable also to flat spacetimes. The method generalizes to more general quantum field theories. The formulae are then applied to a few problems to illustrate their use. 
\end{abstract}

\maketitle

\section{Introduction}

It is well known that calculations in perturbative quantum field theory in curved spacetimes are exceedingly difficult. The propagators are hard to determine, and quantization is impossible for arbitrary spacetimes; particle interpretations can only be established in very few time-dependent spacetimes \cite{Birrell1982}\cite{Wald1994}. The existence of the S-matrix is also dubious, though can be proved in some spacetimes (notably static spacetimes and a restricted class of non-static ones) \cite{Wald1979S}. 

Nevertheless, in some spacetimes, perturbative calculations are possible. Various Robertson-Walker models, important in cosmology, are examples of a metric in which mode functions can be determined \cite{Birrell1979b}. In principle, perturbative calculations can be performed using an Lehmann-Symanzik-Zimmerman\footnote{Typically abbreviated LSZ}-type formalism \cite{Birrell1979}, or by using the Schwinger-Keldysh formalism \cite{Jordan1986}. Calculations can be performed either in the in-out formalism in which the propagators are calculated as the matrix element between two vacuums or the in-in formalism, where one picks either the in or out vacuums and calculates expectation values using only that. The calculations usually are done in the in-in formalism, since it seems to more easily lend itself to physical interpretation \cite{Jordan1986}, and in any case the calculations are often easier \cite{Chen2017}.

Though practical calculations are difficult, there are examples of e.g. scalar decay calculations \cite{Birrell1979b, Lankinen2017, Lankinen2018, Lankinen2018b}, scalar decay in to fermions in Robertson-Walker spacetimes \cite{Lankinen2020}, and solutions for free fermions in assorted spacetimes \cite{Barut1987,Moradi2008}. There are also recent results for backreaction \cite{Taylor2021, Zilberman2021, Meda2021a, Pla2021, Meda2021b, Bernardo2021}. The calculations so far have relied on conformally coupled, massless particles.

Our contribution will be to present the reduction formulae for fermions and gauge fields in globally hyperbolic curved spacetimes, suitable for use in either the in-in or in-out formalisms. Such reduction formulae have previously been published for scalar fields \cite{Birrell1982, Birrell1979}, but to our knowledge not for fermions or gauge fields. We present two ways to derive the reduction formula with the hope that the second way is more technically expedient than what is typically found in textbooks on quantum field theory. We will also apply the formulae to a few topical systems to illustrate their use.

In the following, we will assume that the model under investigation is mathematically well-defined. The relevant criterion is the existence of the S-matrix; for the S-matrix to exist, the Bogoljubov transformation between the different vacua of the theory must exist. This essentially requires that the sum of the Bogoljubov coefficients converges; full details can be found in \cite{Wald1994}. We also require space to be globally hyperbolic and asymptotically stationary, such that it is possible to define vacua for the in and out states; more details in \cite{Birrell1979}. We point out that if the Bogoljubov transform cannot be found, then there is no unitary transformation between the in- and out states and the S-matrix does not exist. In that case, we cannot hope to do any calculations. We therefore limit ourselves to the class of spacetimes in which the S-matrix exists. In particular, it is known to exist for the spacetimes used in \ref{sec:example}.

In the present paper we study the various types of fields in the curved space time, present novel derivations and formulae for spinors 
and gauge fields, and discuss their usage in practical calculations.

\section{Preliminaries}
\subsection{Fermions}

In curved spacetime with metric $g_{\mu\nu}$, the gamma-matrices is modified to satisfy the anti-commutation relations
\begin{align}
    \{\gamma ^\mu , \gamma ^\nu \} = 2 g^{\mu \nu}.
\end{align}
We then define the curved space gamma matrices with the help of the tetrads $e^a_\mu$. The details may be found in \cite{Birrell1982}. Taking advantage of the principle of equivalence, we can set up an inertial coordinate system $\xi $ at every point such that the tetrad is given by its relation to the general curved coordinates:
\begin{align}
    \frac{\partial \xi ^a}{\partial x^\mu }dx^\mu = e^a_{\mu}dx^\mu.
\end{align}
Then, 
\begin{align}
    g_{\mu \nu} &= e^a _\mu e^b _\nu \eta _{ab},\\
    \gamma ^\mu &= e^\mu _{a}\gamma ^{(a)}.
\end{align}
We have adopted the notation that latin indices refer to inertial coordinate systems, and parentheses refer to flat spacetime gamma matrices.

The covariant derivatives $\nabla _\mu $ are then given by
\begin{align}
\nabla _\mu \overline{\psi} (x) &= \partial _\mu\overline{\psi} (x)  - \overline{\psi }(x)\Gamma _\mu , \label{eq:covspinor}\\
\nabla _\mu \psi (x) &= \partial _\mu \psi (x) + \Gamma _\mu \psi (x),\\
		\nabla _\mu \gamma ^\mu &= \partial _\mu \gamma ^\mu +\Gamma ^{\nu}_{\nu \mu }\gamma ^\mu +  [\Gamma _\mu , \gamma ^\mu ] = 0 \label{eq:covgamma}
\end{align}
and the Dirac conjugate is defined as $\overline{\psi} = \psi ^\dagger \gamma ^{(0)}$. Here, the $\Gamma ^{\nu}_{\nu \mu}$ refers to the Christoffel symbol, whereas $\Gamma _\mu$ is the spin connection, explicitly
\begin{align}
    \Gamma _\mu = \frac{1}{8}[\gamma ^{(a)},\gamma ^{(b)}]e_{a}^\nu \nabla _\mu e_{b\nu }.
\end{align}

Quantization proceeds in analogy to the flat spacetime case. One sets up the equal-time anti-commutation relations
\begin{align}
    \{ \psi _a (x,t) , \pi _b (x',t)\} = i\delta _{ab}\delta (x-x'),
\end{align}
where $\pi$ is the conjugate momentum. We then write the field operator as
\begin{align}
    \psi (x,t) = \sum _{s} \sum _k f_s(x,t)b^s_k + g_s(x,t)d^{s\dagger }_k,
\end{align}
where $s$ is the helicity index and $f_s$ and $g_s$ are the mode functions. Further details are available in standard textbooks, for example \cite{Birrell1982}.

With the preceding notations the Dirac Lagrangian in curved spacetime (see e.g. \cite{Birrell1982})
\begin{align}\label{ferL}
    \frac{1}{\sqrt{-g}}\mathcal{L_\psi } = \frac{i}{2}\bigg( \overline{\psi}\gamma ^\mu\nabla _\mu \psi - (\nabla _\mu \overline{\psi})\gamma ^\mu \psi \bigg) - m\overline{\psi}\psi 
\end{align}
 leads to the well-known equation of motion
\begin{align}
    i\gamma ^\mu \nabla _\mu \psi - m\psi = 0.
\end{align}

\subsection{Gauge fields}

The free Lagrangian of a gauge field in curved spacetime is given by
\begin{align}\label{gbL}
    \mathcal{L} &= -\frac{1}{4}\sqrt{-g} \bigg( \nabla ^\mu A^\nu - \nabla ^\nu A^\mu  \bigg) \bigg( \nabla _\mu A_\nu - \nabla _\nu A_\mu \bigg) \nonumber\\
    &= -\frac{1}{4}\sqrt{-g}F^{\mu \nu}F_{\mu \nu}
\end{align}
with $\nabla _\mu$ the usual covariant derivative. This expression of free, quadratic Lagrangian is suitable for all gauge components of any gauge field, commutative or not.
To avoid irrelevant indices, we omit them, or consider only one Abelian gauge field.
Thus the equation of motion of the free gauge field reads as
\begin{align}
    \nabla _\mu F^{\mu \nu} = \nabla _\alpha \nabla ^\alpha  A^\mu - \nabla _\mu \nabla ^\nu A^\mu  = 0 .
\end{align}

Moreover, in this article, we work in the covariant Lorenz gauge $\nabla _\mu A^\mu = 0$, a straightforward generalization of the flat space-time case. We assume that corresponding calculations in some other gauge are also possible, but we have not checked this.

\section{Reduction formulae in curved spacetime}
\subsection{General reduction}\label{sec:genred}

We first deal with the common generic part of the formula, which only depends on the Bogoljubov coefficients and not on the particular type of the field. Note also, that all following calculations should in principle be done using wave packets instead of sharply defined particle states (unnormalized Fock space states), but as in the flat spacetime case, this omission does not cause any difficulties (see almost any standard QFT textbook, e.g. \cite{srednicki_quantum_2007}). So, we have omitted writing out the wave packets to save space and highlight the essential parts of the derivations. Moreover, we have suppressed all irrelevant indices, such as helicity indices, gauge group indices etc. to keep the formulae as simple as possible.

We suppose that $a,a^\dagger$ are the annihilation/creation operators associated with the in-vacuum and $b,b^\dagger$ with the out-vacuum. To these operators are associated mode functions $f$ (in-vacuum) and $g$ (out-vacuum). We suppose, that the initial time is $-\infty$ and the final time is $\infty$. The sets $\{p_i\}$ and $\{k_i\}$ denote the outgoing and incoming four-momenta, respectively. 

Then we are interested in reducing the matrix elements
	\begin{align}
		S = \langle \{p_i\}|\{k_i\}\rangle . \label{eq:Sred}
    \end{align}
to the expectation vacuum values of some well defined operator of the fields.
%Here the sets $\{k_i\}$ and $\{p_i\}$ represents all in and out momenta of the system.
To that we need the Bogoljubov transformations, relation between in- and out-operators, which are given by \cite{Birrell1982}
    \begin{align}
        a_k = \sum _p \bigg( \alpha _{pk}b_p + \beta ^*_{pk}b^\dagger _j \bigg),\label{eq:bogo1}\\
        b_p = \sum _k \bigg( \alpha ^* _{pk}a_k - \beta ^* _{pk}a_k^\dagger \bigg).\label{eq:bogo2}
    \end{align}
 Here the sum should be understood in a generalized sense; if the spectrum is continuous, they may be replaced by integrals. Using the following notation,
    \begin{align}
        a^\dagger _k(\infty) - a^\dagger _k(-\infty) &= F^\dagger[f] \label{eq:srednicki},\\
        a_k(\infty) - a_k(-\infty) &= F[f^*],\label{eq:srednicki2}\\
        b^\dagger _k (\infty ) - b^\dagger _k(-\infty) &= F^\dagger [g],\\
        b_k(\infty ) - b_k (-\infty ) &= F[g^*],
    \end{align}
where $F$ is a hitherto unknown functional of the mode functions \footnote{Functional in the generalized sense: $F$ is a map from functions to operators.}, we are able to write
the matrix element in a reduced form.
 
We express \eqref{eq:Sred} as
	\begin{align}
		\langle \{p_i\}|\{k_i\}\rangle =\langle \{p_i\}| a_k^\dagger(-\infty ) | \{k_i\}\setminus k \rangle ,
	\end{align}
%We now follow the method found in e.g. \cite{srednicki_quantum_2007}, 
and writing from \eqref{eq:srednicki}
 \begin{align}
     a^\dagger_k(-\infty ) = a^\dagger _k (\infty ) -F^\dagger[f]
 \end{align}
%Here we are assuming the existence of the S-matrix as outlined in the introduction. 
we obtain
	\begin{align}
		\langle \{p_i\}| a_k^\dagger(-\infty ) | \{k_i\}\setminus k \rangle = \langle \{p_i\}| a_k^\dagger(\infty) -F^\dagger[f] | \{k_i\}\setminus k \rangle .
	\end{align}
Next we combine the Bogoljubov transformations \eqref{eq:bogo1} and \eqref{eq:bogo2}
	\begin{align}
		a_k^\dagger(\infty)=\sum _{pq} \alpha ^{-1}_{k,p}b^\dagger_p(\infty) + \alpha ^{-1}_{kp}\beta _{pq}a_q(\infty)
	\end{align}
	and apply \eqref{eq:srednicki2} to the latter annihilation operator, with the aim of getting something we can use to operate on the right-side $-\infty$ in-vacuum. We get
	\begin{align}
		&\sum _{pq} \alpha ^{-1}_{k,p}b^\dagger_p(\infty) + \alpha ^{-1}_{kp}\beta _{pq}a_q(\infty) \nonumber\\
  &= \sum _{pq} \alpha ^{-1}_{k,p}b^\dagger_p(\infty) + \alpha ^{-1}_{kp}\beta _{pq}\bigg( a_q(-\infty) + F[f^*] \bigg),
	\end{align}
	which finally leads to the formula having the appropriate annihilation and creation operators:
 \begin{widetext}
	\begin{align}
		&\langle \{p_i\}| a_k(\infty) -F^\dagger[f] | \{k_i\}\setminus k \rangle \nonumber \\
		&= \langle \{p_i\}| -F^\dagger[f]+ \sum _{pq} \alpha ^{-1}_{k,p}b^\dagger_p(\infty) + \alpha ^{-1}_{kp}\beta _{pq}\bigg( a_q(-\infty) + F[f^*] \bigg)| \{k_i\}\setminus k \rangle \nonumber\\
		&=\langle \{p_i\}|-F^\dagger[f]|\{k_i\}\setminus k\rangle + \sum _{p_j \in \{p_i\}}\langle \{p_i\}\setminus p_j| \alpha _{kp_{j}} ^{-1}|\{k_i\}\setminus k\rangle \nonumber\\
  &+ \sum _{q\in \{k_i\}\setminus k} \langle \{p_i\}| \alpha ^{-1}_{k,p}\beta _{p,q}|\{k_i\}\setminus \{k,q\}\rangle + \langle \{p_i\}| \sum_{pq} \alpha ^{-1}_{kp}\beta _{pq} F[f^*]|\{k_i\}\setminus k\rangle .\label{eq:rightred}
	\end{align}
 \end{widetext}
 This is the single particle reduction from the right.
	
The same steps using
	\begin{align}
		a_k(-\infty) &= \sum _p (\alpha _{pk} b_p(-\infty) + \beta _{pk}^* b^\dagger_p (-\infty)) \nonumber\\
         &\Leftrightarrow\nonumber\\
        b_p(-\infty) &= \sum _k \alpha ^{-1}_{kp}a_k(-\infty) - \sum _{qk}\alpha ^{-1}_{qp}\beta ^* _{kq}b^\dagger_k(-\infty ),
	\end{align}
	give immediately single particle reduction from left:

	\begin{align}
		&\langle \{p_i\}\setminus p|b_p(\infty)|\{k_i\}\rangle \nonumber\\
  &=\sum _{k\in \{k_i\}}\langle \{p_i\}\setminus p|\alpha ^{-1}_{kp}|\{k_i\}\setminus k\rangle  \nonumber\\
  &-\sum _q \sum _{p_j\in \{p_i\}\setminus p}\langle \{p_i\}\setminus \{p,p_j\}|\alpha ^{-1}_{qp}\beta ^*_{p_jq}|\{k_i\}\rangle\nonumber\\
  &-\sum _{qk}\langle {p_i}\setminus p|\alpha ^{-1}_{qp}\beta ^*_{kq}F^\dagger[g]|\{k_i\}\rangle + \langle \{p_i\}\setminus p|F[g^*]|\{k_i\}\rangle . \label{eq:leftred}
	\end{align}

	In the corresponding in-in calculation, the difference is that the parts of \eqref{eq:leftred} and \eqref{eq:rightred} dependent on the Bogoljubov coefficients vanish as the in- and out vacuums are the same. 
 
 To get the desired form of the reduction formula we need to find the functional $F[f]$. For scalars $F$ is well-known and it is given by 
 \begin{align}
     F^\dagger[f] = -i\int d^4x \sqrt{-g}\, w(x)^* K_x \phi (x),
 \end{align}
 where $K_x=\Box + m^2$ is the curved spacetime Klein-Gordon operator with the curved spacetime D'Alembertian $\Box$  \cite{Birrell1979}, and $w$ is the positive energy mode function. Next we derive the reduction formulae for spin $\frac 12 $ and spin 1 fields.
 
     \subsection{The fermion formula}
     \subsubsection{Derivation by EOM manipulation}\label{sec:direcmet}
     
The reduction formula for fermions is not previously found in the literature. We derive it in two ways. First by brute force, direct calculation commonly found most textbooks. 
%(such as \cite{srednicki_quantum_2007}) 
Secondly using conserved currents. The comparison between the two methods emphasises the relative simplicity of the latter method.

The Lagrangian \eqref{ferL} leads to
% \begin{align}
 %       \frac{1}{\sqrt{-g}}\mathcal{L_\psi } = \frac{i}{2}\bigg( %\overline{\psi}\gamma ^\mu\nabla _\mu \psi - (\nabla _\mu %\overline{\psi})\gamma ^\mu \psi \bigg) - m\overline{\psi}\psi . %\label{eq:lagrangianfermion}
 %   \end{align}
the equation of motion for a massless fermion given by
	\begin{align}
		\nabla _\mu \overline{\psi} (x) \gamma ^\mu &= 0 \label{eq:eom}
	\end{align}
with the covariant derivative satisfying equations \eqref{eq:covspinor} and \eqref{eq:covgamma}. The inclusion of (Dirac) mass does not change the final formula, but
 %We use the massless fermion because the addition of mass would 
 merely make the intermediate expressions lengthier; it will be evident by the end that there is no essential difference in the calculation. We choose the inner product \footnote{The use of the term inner product is merely conventional. Equation \eqref{eq:innerprod} does not fulfill the typical criteria of an inner product.}
    \begin{align}
		\langle \varphi | \psi \rangle = \int d^3x \sqrt{-g}\,\overline{\varphi}(x) \gamma ^0 \psi (x) \label{eq:innerprod}
	\end{align}   
and we are looking for the difference \eqref{eq:srednicki} between the creation/annihilation operators in the far past and future.
 
 When the field $\psi$ has a mode expansion \footnote{As usual, the general ladder operators $a_k$ and $b_k$ are replaced by their fermionic counterparts $b_k^s$ and $d_k^s$, respectively.} 
    \begin{align}
		\psi (x) = \sum _{s}\sum _k u_k^s(x) b^s_k + v_k^s(x)d^{s\dagger }_k\,
	\end{align}
respect to an orthonormal basis of modes $u_k^s(x)$ (and $v_k^s(x)$) satisfying the equation of motion \eqref{eq:eom} we have
	\begin{align}
		b_k^s(t) = \langle u_k^s | \psi \rangle = \int d^3x \sqrt{-g}\, \overline{u}_k^s(x) \gamma ^0 \psi (x), \label{eq:annihdef}
	\end{align} 
when the mode functions are orthonormal with respect to the inner product \eqref{eq:innerprod}. We will henceforth suppress the helicity indices $s$, because they are of no consequence for the following calculation.

 Let us first use the fundamental theorem of calculus in operator form:
	\begin{align}
		b_k(\infty) - b_k(-\infty) = \int _{-\infty}^\infty dt\, \partial _0 b_k(t). \label{eq:diffop}
	\end{align}
Combining it and \eqref{eq:annihdef}, we get 	
	\begin{align}
		&b_k(\infty) - b_k(-\infty) \nonumber\\
        &= \int d^4x \partial_0 \left (\sqrt{-g}\,\overline{u}_k(x) \gamma ^0 \psi (x)\right ) \nonumber\\
		&= \int d^4x\sqrt{-g} \bigg[ \frac{1}{\sqrt{-g}}(\partial_0 \sqrt{-g})\overline{u}_k(x) \gamma ^0 \psi (x) + (\partial_0\overline{u}_k(x)) \gamma ^0 \psi (x) \nonumber\\
		&+ \overline{u}_k(x) (\partial_0\gamma ^0) \psi (x) + \overline{u}_k(x) \gamma ^0 (\partial_0\psi (x))\bigg]. \label{eq:fulldiff}
	\end{align}
From the equation of motion \eqref{eq:eom} we get
	\begin{align}
		\partial _0 \overline{u}_k(x)\gamma ^0 = -\nabla _i\overline{u}_k(x) \gamma ^i + \overline{u}_k(x)\Gamma _0 \gamma ^0.
	\end{align}
We insert this to the second term in the square brackets of the eq. \eqref{eq:fulldiff}, obtaining
	\begin{align}
		&(\partial_0\overline{u}_k(x)) \gamma ^0 \psi (x) \nonumber\\
        &=  \bigg(-\nabla _i\overline{u}_k(x) \gamma ^i + \overline{u}_k(x)\Gamma _0 \gamma ^0\bigg) \psi (x) \nonumber\\
		&= \bigg(-\partial  _i\overline{u}_k(x)\gamma ^i  + \overline{u}_k(x)\Gamma _i\gamma ^i + \overline{u}_k(x)\Gamma _0 \gamma ^0\bigg) \psi (x).
	\end{align}
Taking into account the integration of the term it can be rewritten as
  \begin{widetext}
	\begin{align}
		&\int d^4x\sqrt{-g}(\partial_0\overline{u}_k(x)) \gamma ^0 \psi (x) \nonumber \\
		&= \int d^4x\sqrt{-g}\bigg[  \bigg(-\partial  _i\overline{u}_k(x)\gamma ^i  + \overline{u}_k(x)\Gamma _i\gamma ^i + \overline{u}_k(x)\Gamma _0 \gamma ^0\bigg) \psi (x) \bigg] \nonumber\\
		&= \int d^4x \sqrt{-g}\bigg[ \bigg(\overline{u}_k(x)\partial  _i\gamma ^i + \frac{1}{\sqrt{-g}}  \overline{u}_k(x)\gamma ^i \partial _i \sqrt{-g} +\overline{u}_k(x)\Gamma _i\gamma ^i + \overline{u}_k(x)\Gamma _0 \gamma ^0\bigg) \psi (x) + \overline{u}_k(x) \gamma ^i\partial _i \psi (x)\bigg]\nonumber\\
		&=\int d^4x \sqrt{-g}\bigg[ \bigg(\overline{u}_k(x)\partial  _i\gamma ^i + \frac{1}{\sqrt{-g}}  \overline{u}_k(x)\gamma ^i \partial _i \sqrt{-g} +\overline{u}_k(x)\Gamma _\mu \gamma ^\mu \bigg) \psi (x) + \overline{u}_k(x) \gamma ^i\partial _i \psi (x)\bigg].
	\end{align}
 \end{widetext}
Where the second and third equalities follow from simply taking the derivatives and applying Gauss' theorem. Inserting the result in to eq. \eqref{eq:fulldiff} we get
  \begin{widetext}
	\begin{align}
        & b_k(\infty) - b_k(-\infty) \nonumber\\
		&= \int d^4x\sqrt{-g} \bigg[ \frac{1}{\sqrt{-g}}(\partial_0 \sqrt{-g})\overline{u}_k(x) \gamma ^0 \psi (x) + (\partial_0\overline{u}_k(x)) \gamma ^0 \psi (x) \nonumber\\
		&+ \overline{u}_k(x) (\partial_0\gamma ^0) \psi (x) + \overline{u}_k(x) \gamma ^0 (\partial_0\psi (x))\bigg] \nonumber\\
		&= \int d^4x \sqrt{-g}\bigg[ \overline{u}_k(x)\bigg(\partial _\mu \gamma ^\mu + \Gamma _\mu \gamma ^\mu \bigg)  \psi (x) + \overline{u}_k(x)\frac{1}{\sqrt{-g}}\gamma ^\mu \partial _\mu  \sqrt{-g} \psi(x) + \overline{u}_k(x)\gamma ^\mu \partial _\mu \psi (x) \bigg]. \label{eq:laststep}
	\end{align}
  \end{widetext}
Now we can use eq. \eqref{eq:covgamma} to modify the first term in this expression:
	\begin{align}\label{midr}
		&\overline{u}_k(x)\bigg(\partial _\mu \gamma ^\mu + \Gamma _\mu \gamma ^\mu \bigg)  \psi (x) \nonumber\\
    &= \overline{u}_k(x)\bigg(-\Gamma ^\nu _{\nu \mu}\gamma ^\mu  + \gamma ^\mu \Gamma _\mu \bigg)  \psi (x) \nonumber\\
		&=\overline{u}_k(x)\bigg(-\frac{1}{\sqrt{-g}}\partial _\mu \sqrt{-g}\gamma ^\mu   + \gamma ^\mu \Gamma _\mu \bigg)  \psi (x).
	\end{align}
	
 Finally, using \eqref{midr} in eq. \eqref{eq:laststep}
	\begin{align}
		b_k(\infty) - b_k(-\infty) = \int d^4x\sqrt{-g}\,\overline{u}_k(x) \gamma ^\mu \nabla _\mu \psi (x). \label{eq:lszdiff}
	\end{align} 
	 This result generalises the corresponding flat space formula, as expected. Using the massive fermion field
  with a mass $m$, we get the result 
\begin{align}
    b_k^s(\infty) - b_k^s(-\infty) = \int d^4x\sqrt{-g}\,\overline{u}_k^s(x) (i\gamma ^\mu \nabla _\mu - m) \psi (x),
\end{align}
 where we have replaced the omitted helicity indices: the derivation is completely independent of them. The corresponding calculation for $d_k^s$ operators follows the same pattern.
 
  \subsubsection{Using a conserved current}
  
Let us now introduce the conserved current method as a way of finding reduction formulae using conserved currents. The idea is to find a conserved current and use it to find an inner product with respect to which the mode functions are orthogonal. This inner product -- taken as the zeroth component of the conserved current -- is then used to derive the reduction formula in a simple manner. The reference \cite{mostafazadeh_quantum_2006} provides further details on the requirements of inner products in relativistic quantum mechanics.

We now find the appropriate inner product with respect to which the mode functions are orthogonal. In curved spacetimes, we should keep in mind that the words "appropriate inner product"\ do quite a bit of heavy lifting: inner products in QFT are not unique, since it is possible to explicitly construct unitarily inequivalent inner products \cite{Wald1994}. Unitarily equivalent inner products give the same physical results, but inequivalent ones may not; the inner product has to be fixed by some other method, like experimental data or defining the inner product to have the appropriate flat spacetime limit. Finding an appropriate structure might be difficult in the most general case, but it is possible for fermions and gauge fields.
     
We find an appropriate inner product by considering the global and infinitesimal phase transformation $\psi\rightarrow \psi ' = e^{i\chi }\psi\approx (1+i\chi )\psi $ and using the the inner product determined by the conserved Noether charge.
%, just as is done in the flat spacetime case. 
This guarantees that the inner product is conserved in time, independent on the time-slice, and thus allowing the sotr of probability interpretation mentioned in \cite{mostafazadeh_quantum_2006}. In the following, we assume that the spacetime is globally hyperbolic and that it is possible to choose appropriate coordinates such that $x^0$ marks the time direction. Any other coordinate system would work, but this is the most convenient one.
	
Starting directly from a standard expression for the change of the action
	\begin{align}
		\Delta S = \int_V d^4x \bigg( \sum _a &\bigg[ \frac{\partial \mathcal{L}}{\partial \psi _a} - \partial _\mu \frac{\partial \mathcal{L}}{\partial (\partial _\mu \psi _a)}\bigg]\Delta \psi _a + \nonumber\\
        \partial _\mu \sum _a &\bigg[\frac{\partial \mathcal{L}}{\partial (\partial _\mu \psi _a)}\Delta \psi _a   \bigg]\bigg) = 0, \label{eq:change}
	\end{align}
where $V$ is a space-time volume bounded by two space-like surfaces: $\partial V= \sigma_1 \cup \sigma_2$.
The last equality is due to the global U(1) symmetry of the Lagrangian (both interacting and non-interacting, for the interactions considered here). The index $a$ here runs over the components of the fermion field and its Dirac conjugate.
%, and we've bundled the spin indices in to $a$. 

We can calculate the current (second bracketed term) directly:
	\begin{align}
	\int d^4 x\, \partial _\mu \sum _a &\bigg[\frac{\partial \mathcal{L}}{\partial (\partial _\mu \psi_a)}\Delta \psi _a\bigg] 	\nonumber \\
 &=\int d^4 x\, \partial _\mu \bigg[\sqrt{-g}\,\overline{\psi}(x) \gamma ^\mu \psi (x)  \bigg]\chi  \label{eq:conscurrent}
	\end{align} 
and if $\psi$ and $\overline{\psi}$ are solutions of their equation of motion, then the first bracketed term in \eqref{eq:change} is zero. Applying Gauss' theorem as usual to \eqref{eq:conscurrent} and using \eqref{eq:change} we get
\begin{widetext}
	   \begin{align}
	   0&= \int _{\sigma _1} d^3 x\sqrt{-g}\, n_\mu \bigg[\overline{\psi}(x) \gamma ^\mu \psi (x)  \bigg]\chi - \int _{\sigma _2}d^3 x\sqrt{-g}\,n_\mu\bigg[\overline{\psi}(x) \gamma ^\mu \psi (x)  \bigg]\chi \\
		&\iff  \nonumber\\
        0&= \int _{\sigma _1} d^3 x\sqrt{-g}\bigg[\overline{\psi}(x) \gamma ^0 \psi (x)  \bigg] - \int _{\sigma _2}d^3x\sqrt{-g}    \bigg[\overline{\psi}(x) \gamma ^0 \psi (x)  \bigg].
	    \end{align}
\end{widetext}
Here $n_\mu$ is a future-oriented unit vector, while the spacelike surfaces $\sigma _i$ define a foliation of the spacetime. For the second line we have used our special coordinate system and the fact that the space is globally hyperbolic and $\chi\ne 0$. This quantity is clearly conserved in time.
    
Now we  can replace the field appearing in the Dirac conjugate $\overline{\psi}(x)$ by completely independent 
spinor field $\psi'$ which has same gauge transformation  and obeys the same equation of motion as $\psi$.
The current that we get is equally well conserved and has the form
\begin{equation}
J^\mu[\psi ',\psi] = \overline{\psi'}\gamma ^\mu \psi.
\end{equation}
This gives the general form of the inner product determined by the phase transformation, eq. \eqref{eq:innerprod}.

Replacing the field $\psi'$ by a mode $u_k$, using Gauss' theorem and knowing that $u_k$ is a solution of the equation of motion but $\psi$ is interacting,
    \begin{align}
        &\int d^4x \sqrt{-g}\nabla _\mu J^\mu[u_k,\psi] \nonumber \\
        &= \left (\int _{\sigma_2}-\int _{\sigma_1}\right )\sqrt{-g}\,J^0[u_k,\psi ] d^3x \nonumber\\&=b_k(\infty)-b_k(-\infty), \label{eq:lszferm}
    \end{align}
where  $\sigma _1$ and $\sigma _2$ are the constant time surfaces, which are set to limits $t\rightarrow \pm \infty$, correspondingly. The first equality follows from Gauss' theorem and the second from \eqref{eq:annihdef}.
The equation \eqref{eq:lszferm} provides the functional $F$ which can then be plugged in to equations \eqref{eq:rightred} and \eqref{eq:leftred}, thus providing us with a reduction formula.
	
The latter procedure is completely general: if a conserved current is available as a sesquilinear inner product, we can use it to derive a reduction formula. Even if the current is not of the Noetherian, we can still use the procedure, as we will see presently.

The inner product we have used here is by no means unique, which is a typical situation in quantum field theory. Let us consider an example of another current we could conceivably use for an inner product in the massless fermion case. The transformation $\psi \mapsto \psi' =e^{i\chi \gamma ^\mu \gamma ^5}\psi$ (with curved $\gamma ^\mu$ and $\gamma ^5$) generates another conserved Noether charge and inner product, and leads to the functional formula
    \begin{align}
        F[u^*]=\int d^4x \sqrt{-g}\, \overline{u}^s_k(x) \gamma^5 \gamma^0 \psi (x).
    \end{align}
This is the same functional as previously except with a redefinition $u \mapsto \gamma ^5 u$. If $u$ satisfies the EOM, then so does $\gamma ^5 u$ as can be easily checked:
    \begin{align}
        \gamma ^\mu \nabla _\mu u_k = 0 \iff 0= \gamma ^5 \gamma ^\mu \nabla _\mu u_k  =\gamma ^\mu \nabla _\mu (\gamma ^5 u_k),
    \end{align}
 where in the last equality we used the covariance properties of the curved spacetime $\gamma ^5$. 

The operators defined with this inner product are evidently not quite the same as the ones in the foregoing calculation, but nevertheless they are clearly unitarily equivalent. For massless fermions, there does not seem to be an obvious reason to prefer one over the other.

A word of warning about using conserved currents in deriving reduction formulae is in order: it relies on assuming formula \eqref{eq:annihdef} is valid even when $\psi$ is interacting and lives in different Hilbert space that the non-interacting field. This is a standard assumption made in QFT textbooks like \cite{srednicki_quantum_2007}, but runs afoul of Haag's theorem \cite{haag_quantum_1955}. In flat spacetime, this difficulty is not considered serious, since Haag's theorem relies on idealized assumptions that are presumably not realized in a practical calculation. In curved spacetime, the additional difficulty is that the S-matrix may not exist in some spacetimes -- such spacetimes cannot be used for scattering calculations, so we do not concern ourselves with them. The foregoing calculation or basically any calculation using both interaction and non-interacting fields is strictly speaking non-rigorous. It works in the same sense as the standard calculations work: as a mnemonic that can be made more rigorous by careful study, in particular when using perturbation theory.

\subsection{The gauge field formula}
    
The equation of motion for a real spin-1 gauge field in curved spacetime derived from eq. \eqref{gbL} is
    \begin{align}
        \nabla _\mu F^{\mu \nu}[A] = 0, \label{eq:veceom}
    \end{align}
where $\nabla _\mu$ is the curved spacetime derivative and we have denoted
    \begin{align}
        F^{\mu \nu}[A] = \nabla ^\mu A^\nu - \nabla ^\nu A^\mu .
    \end{align}
As discussed earlier, we use the Lorenz gauge with $\nabla _\mu A^\mu = 0$. The choice is based on convenience alone.
    
First of all, we establish that the following current is conserved when $A,A'$ are solutions of \eqref{eq:veceom} in covariant Lorentz gauge and thus can be used as an inner product:
    \begin{align}
        J_\nu[A',A] = -iA'_\mu \overset{\leftrightarrow}{\nabla}_\nu  A^\mu .
    \end{align}
    Let us first collect some general formulas to be used: 
    \begin{align}
         \nabla _\mu \nabla^\mu A^\nu &= \nabla _\mu \nabla^\nu A^\mu + \nabla _\mu  F^{\mu \nu}, \label{eq:vecident1}\\
        [\nabla _\mu , \nabla _\nu]A^\alpha &= R^\alpha _{\ \ \delta \nu \mu} A^\delta, \label{eq:vecident2}\\
        B_\mu  \nabla_\nu \nabla^\mu A^\nu &= B_\mu [\nabla _\nu, \nabla ^\mu] A^\nu + B_\mu \nabla ^\mu \nabla _\nu A^\nu \nonumber\\
        &= B_\mu [\nabla _\nu, \nabla ^\mu] A^\nu. \label{eq:vecident3}
    \end{align}
The first formula is only the equation of motion, the second eq. is well-known commutator of covariant derivatives, and the third uses the gauge condition.  

As in the case of the scalar field, the use of the inner product relies on an extension of the real field to the complex space. When $A^\mu$ satisfies \eqref{eq:veceom}, the latter term in \eqref{eq:vecident1} is zero. The complexification is needed to enable the inner product for the positive/negative energy modes. While assuming $A'=B$ does not satisfy \eqref{eq:veceom} but $A^\mu$ does, we have
    \begin{align}
        &\nabla _\nu J^\nu \nonumber\\
        &= B_\mu \nabla _\nu \nabla ^\nu A^\mu - (\nabla_\nu \nabla^\nu B_\mu) A^\mu \\
        &= B_\mu \nabla_\nu \nabla^\nu A^\mu - (\nabla^\nu \nabla_\mu B_\nu) A^\mu - (\nabla _\nu F[B]^\nu _{\ \ \alpha})  A^\alpha \\
        &= B^\mu [\nabla_\nu , \nabla_\mu] A^\nu - ([\nabla_\nu , \nabla_\mu]B^\nu) A^\mu - (\nabla _\nu F[B]^\nu _{\ \ \alpha}) A^\alpha\\
        &= B^\mu R^\nu _{\ \ \delta \mu \nu}A^\delta - R^{\nu}_{\ \ \delta \mu \nu}B^\delta A^\mu - (\nabla _\nu F[B]^{\nu}_{\ \ \alpha})A^\alpha \\
        &= -(\nabla _\nu F^\nu _{\ \ \alpha}[B]) A^\alpha
    \end{align}
employing eqs. \eqref{eq:vecident1}, \eqref{eq:vecident3}, and \eqref{eq:vecident2} in a row. The last expression is clearly identically 0 if $B$ satisfies the EOM as well, so this is an appropriate current for an inner product. As in the previous section, we will assume that $B$ does not satisfy the free equation of motion, since it is supposed to be interacting.

We may now directly apply the ideas of the previous section. We write
     \begin{align}
        a_k^\dagger (t)  = \langle u | A\rangle &= \int_t d^3 x\, \sqrt{-g}\, J^0[u,A] \nonumber\\
        &= \int _t d^3x \sqrt{-g}\, u_{k\alpha}^*  \overset{\leftrightarrow\ \,}{\nabla^0} A^\alpha ,
    \end{align}
where $u_\mu$ is the mode function associated to ladder operator $a_k$.
Then, if $A$ is an interacting field and $u$ a free mode function, be Gauss' theorem we get
    \begin{align}
        \int d^4x \bigg[\sqrt{-g}\, \nabla _\mu J^\mu[u,A] \bigg]& \nonumber\\
        = \int_{\sigma} d^3x \sqrt{-g}\,J^0[u,A] 
        &= a_k^\dagger (\sigma _2)-a_k^\dagger(\sigma _1)
    \end{align}
 where $\sigma = \sigma _2 \cup \sigma _1$ is defined as previously. We then take the limit as $\sigma _i \rightarrow \pm\infty$:
 \begin{align}
     a_k^\dagger(-\infty)-a_k^\dagger(\infty ) &= -i\int d^4 x \sqrt{-g}\nabla _\mu F^\mu _{\ \ \alpha}[A] (u_k^\alpha)^*. \label{redgb}
 \end{align}
 Hence we have derived the reduction formula.

A note about the nature of the vector field is in order. The simple non-gauge vector field with Lagrangian ${\cal L'} = \frac{1}{2}\sqrt{-g}\, \nabla_\mu A_\nu \nabla^\mu A^\nu$ leads to the very same reduction formula \eqref{redgb} as in the case of gauge bosons, but without the need of the gauge condition. The only difference is that the equation of motion  $\nabla _\mu F^\mu _{\ \ \alpha}[A]$ is replaced by $\nabla_\mu \nabla^\mu A_\alpha$.

\section{Using the formulae}\label{sec:example}
\subsection{The general setup}
    
To summarize, we first utilize the the general reduction formulae in section \ref{sec:genred} until we have vacuums on both sides of the bra-ket. We then plug in the calculated functionals $F[g]$, which we put into the table \ref{table} for convenience. The one-particle reduction formulae given above can be used recursively for any number of particle in/out states.    
%(for fermions, there are two sets of operators and for vectors two there are polarization directions; %the changes needed to find all of them are obvious):
    \begin{table*}\label{table}
    \centering
    \renewcommand*{\arraystretch}{1.6}
    \begin{tabular}{c|c|c}
                 & $F[f^*]$ & $F^\dagger [f]$ \\ \hline
     Scalars     &   $-i\int d^4x \sqrt{-g}\, u^*(x) K_x \phi (x) $ & $-i\int d^4x \sqrt{-g}\, u(x) K_x \phi (x)$ \\  \hline 
     Fermions    & $-i\int d^4x \sqrt{-g}\, \overline{u}^s_k(x) D_x \psi (x) $ & $-i\int d^4x \sqrt{-g}\,  \overline{\psi} (x)\overset{\leftarrow}{\overline{D}}_x u^s_k(x)$  \\ \hline
     Vectors     & $-i\int d^4x \sqrt{-g}\, \nabla _\mu F^\mu _{\ \ \alpha}[A] u^{*\alpha}(x)$&  $-i\int d^4x \sqrt{-g}\, \nabla _\mu F^\mu _{\ \ \alpha}[A] u^\alpha (x)$\\ 
    \end{tabular}
    \caption{Functionals $F$ for all the known cases with $K_x = \Box + m^2$ the curved spacetime Klein-Gordon operator and $D_x=i\gamma ^\mu \nabla _\mu - m$ and $\overline{D}_x = i\gamma ^\mu \nabla_\mu + m$ the Dirac curved spacetime Dirac operators \cite{Birrell1982}. For fermions, there are two sets of operators, and for vectors, two polarization directions; adding those indices happens as expected.}
    \end{table*}
    
We will utilize these formulae for two already known cases to illustrate how they work: the scalar decay in to fermions reported in \cite{Lankinen2020} and the classic Bogoljubov coefficient calculation in \cite{Birrell1982, dewitt_quantum_1975}.
    
%Let us first utilize this in the case of an in-in calculation, in which case the general part of the %formula is trivial (both vacuums are the same so there is no Bogoljubov transformation).

\subsection{In-in calculation: Decay of a scalar field}

We now apply the reduction formula in the in-in formalism to the decay of a massive scalar particle in to massless fermions. There is only a single vacuum in the calculation so there is no need for a Bogoljubov transformation.

Let us denote the fermion (Dirac) operator at point $x$ by $D_x$ and the Klein-Gordon operator as $K_x$. The propagators are normalized as
    \begin{align}
        D_x G^{D}_f (x-x') = -\frac{1}{\sqrt{-g_x}}\delta (x-x') \label{eq:gnnorm}
    \end{align}
with a corresponding normalization for the Klein-Gordon operator.
    
We are looking for the scattering element $\langle 1^\psi _{k_1}1^{\overline{\psi}}_{k_2}|1^\phi _p\rangle$. Now
    \begin{align}
        \psi (t,x) &= \sum _s \int d^3k \bigg[ b^s_ku^s_k(t,x) + d^{s\dagger}_kv^s_k(t,x)  \bigg] \\
        \overline{\psi} (t,x) &= \sum _s \int d^3k \bigg[ d^s_k\overline{v}^s_k(t,x) + b^{s\dagger}_k\overline{u}^s_k(t,x) \bigg] \\
        \phi (t,x) &= \int d^3k \bigg[a_kw(t,x) + c_k^\dagger w^*(t,x) \bigg]
    \end{align}
with $b,d$ and $a,c$ the fermionic and bosonic annihilation operators, respectively, satisfying the usual (anti)commutation relations.

    We then use the reduction formulae. Since the in- and out-vacuums are exactly the same, $\alpha _{mn}=\delta _{mn}$ and $\beta_{nm} = 0$. Then, using \eqref{eq:leftred} and \eqref{eq:rightred}, the only term left is the one with no Bogoljubov coefficients. Note that $\langle k_i|k_j\rangle = \delta _{ij}$, which is why the term with only $\alpha _{nm}$ is zero. We get
    \begin{align}
        &\langle 1^\psi _{k_1}1^{\overline{\psi}}_{k_2}|1^\phi _p\rangle \nonumber\\
        &= -\langle 0 | F_{\text{fermion}}[\overline{u}]F_{\text{fermion}}[v] F_{\text{scalar}}[w]|0\rangle\nonumber \\
        &= \int dy_1\sqrt{g_{y_1}} \int dy_2\sqrt{g_{y_2}} \int dx_1\sqrt{g_{x_1}}\overline{u}^s_k(t,x) \nonumber\\
       &\times v^{s'}_k(t,x)w(t,x) D_{y_1}D_{y_2}K_{x_1}\langle 0| T\psi (y_1) \overline{\psi}(y_2) \phi (x_1) |0\rangle \label{eq:reduced}
    \end{align}
    where $D_{y_i}$ are the Dirac operators and $K_{x_1}$ is the Klein-Gordon operator.

    Let us specify the interaction as
    \begin{align}
        \mathcal{L}_{int} &= T\exp \bigg(-i\lambda \int dz \sqrt{-g_z}\ \overline{\psi}(z)\phi(z)\psi(z)  \bigg) \\
        &\approx 1 - i\lambda \int d_z \sqrt{g_z}\ \overline{\psi}_0(z)\phi_0 (z) \psi_0 (z)
    \end{align}
    with the subscript 0 indicating a free field.
    Then the field operator expectation value in \eqref{eq:reduced} is written as
    \begin{align}
        &\langle 0| T\psi (y_1) \overline{\psi}(y_2) \phi (x_1) |0\rangle \nonumber\\
        &= - i\lambda \langle 0| T\psi _0(y_1) \overline{\psi}_0(y_2) \phi _0(x_1)\nonumber\\
        &\times\int dz \sqrt{g_z}\ \overline{\psi}_0(z)\phi_0 (z) \psi_0 (z)|0\rangle .\label{eq:firstorder}
    \end{align}

    Using Wick's theorem, we can write \eqref{eq:firstorder} as
    \begin{align}
        &\langle 0| T\psi (y_1) \overline{\psi}(y_2) \phi (x_1) |0\rangle \nonumber\\
        &= -i\lambda G^D_f(y_1-z)G^D_f(y_2-z) G^K_f(x-z) .\label{eq:firstorderwithgreen}
    \end{align}
    Applying the Dirac and Klein-Gordon operators to this expression and using the normalization \eqref{eq:gnnorm}, we get immediately
    \begin{align}
        -iD_{y_1}&D_{y_2}K_{x}\lambda G^D_f(y_1-z)G^D_f(y_2-z) G^K_f(x-z) \\
        &= i\frac{1}{\sqrt{-g_{y_1}g_{y_2}g_{x}}}\bigg[\delta (y_1-z)\delta (y_2-z) \delta (x-z)\bigg]. \label{eq:diracs}
    \end{align}
    Inserting this to \eqref{eq:reduced}, we get after a few integrations
    \begin{align}
        \langle 1^\psi _{k_1}1^{\overline{\psi}}_{k_2}|1^\phi _p\rangle  = i\lambda \int dz \sqrt{-g}\bigg[\overline{u}^s_k(t,x)v^{s'}_k(t,x)w(t,x)\bigg] .
    \end{align}
    
    \subsection{In-out calculations: using Bogoljubov coefficients}
    
    Let us then deal with an example where we use only the Bogoljubov coefficients. For concreteness, starting with a 0-particle state in the Robertson-Walker metric, we wish to end up with an 2-particle state due to space-time particle creation. We assume there are no interactions, so that only spacetime particle creation is relevant. We deal with the scalar case to avoid the cumbersome use of indices. Then using formula \eqref{eq:leftred} we get
    \begin{align}
        \langle 1^\phi_{k_1} 1^\phi _{k_2}|0\rangle = -\sum _p \langle 0 |\alpha ^{-1}_{pk_1}\beta ^* _{k_2 p} |0\rangle ,
    \end{align}
    and furthermore
    \begin{align}
        \langle  \{ 1^\phi _{k_i} \} |0\rangle = 0 \quad  \text{if}\ \  |\{ 1^\phi _{k_i} \}| \equiv 0 \mod 2
    \end{align}
	with the vertical bars indicating the cardinality of the set. We see the expected result: particles are created in pairs. If we had only one particle in the out-state, the amplitude would be zero. If there were more particles to reduce, the formula would then be applied recursively until we end up with a vacuum. If interactions were present, more terms from e.g. \eqref{eq:leftred} would be added; if there were fermions, you would take the formula \eqref{eq:leftred} with the fermionic creation operations, and so on.

    Note that this is a separate issue from \textit{observing} particles. Particles may of course be observed one at a time. Mathematically the construction for a "measurement device"\ is given in standard works, such as \cite{Birrell1982}. We are here dealing with only the amplitudes of particle creation; observation requires a separate treatment.

    \section{Discussion}\label{sec:discussion}

    We have derived curved spacetime reduction formulae for vectors and fermions in arbitrary spacetimes and applied them to a few example problems. In addition, we expressed the Bogoljubov-dependent part of the reduction formulae in a way which we hope is less oblique than the scalar field calculations in \cite{Birrell1979}\footnote{There are also some typographical errors in \cite{Birrell1979} which one should be wary of when applying the formulae.}, and therefore be more practical. The formulae were derived using a method not typically seen in the literature. We hope to use our the reduction formulae for scattering calculations in e.g. cosmological spacetimes in the future.

    The reason for our interest in reduction formulae is that they are relatively formalism-agnostic. Just as is the case in flat spacetime, the reduction formula binds together a variety of formalisms for doing scattering calculations: you can first use the formula and then use whichever method seems suitable to get the vacuum expectation values. We also have not seen these curved spacetime formulas published in full, though \cite{Birrell1982} mentions them and \cite{Birrell1979} includes the full scalar reduction formula. We hope they facilitate more calculations such as those in \cite{Lankinen2017, Lankinen2018, Lankinen2018b, Lankinen2020}.

    In flat spacetimes propagators are in practice obtained either by the operator formalism or by using path integrals. In curved spacetimes, however, there are a variety of methods; for example, variations of Schwinger's method \cite{Jordan1986, Chen2017}, the added-up method as used in \cite{audretsch_improved_1987, audretsch_mutually_1985}, operator methods like those in \cite{Birrell1979}, and even Schrödinger picture methods \cite{long_schrodinger_1996}. The variety of methods is a result of the complications in dealing with spacetime in quantum field theory, but whenever a scattering calculation is at hand, the preceding formulae may be used.

    We used the conserved current method for deriving the reduction formulae in this paper. We have not seen other examples of this method in the literature -- possibly because in the flat spacetime case, it is relatively straightforward to derive the reduction formulae by manipulating equations of motion directly, such as we did here for the fermions. In curved spacetimes, as can be seen in the fermion calculation, this method quickly becomes laborious and technically challenging. The use of the conserved currents not only simplifies the calculation but conceptually relates conservation laws to the inner products used in QFT. It also generalizes to any field theory with conserved currents easily.

    We emphasize that the inner products used in QFT are not fixed by the algebraic structure of the theory. You can explicitly construct unitarily inequivalent representations of the commutation relations \cite{Wald1994}; Haag's theorem also shows that interacting theories are not unitarily equivalent to free ones \cite{haag_quantum_1955}. This allows us to use our conserved current of choice; there may well be other unitarily equivalent choices, as we showed. Yet other choices may well be unitarily inequivalent, and there is no telling if they would lead to the same predictions.

     We think that using conserved currents directly to find the reduction formula might be more transparent and pedagogical than manipulating the equation of motion in the brute-force way, 
     %of e.g. Srednicki \cite{srednicki_quantum_2007}
     since the procedure seems more generally applicable. It also makes the arbitrariness of the inner product obvious, and in our opinion makes clearer the assumptions going in to putting interacting fields in to the current to get the creation operators. At any rate, obtaining reduction formulae using the direct method as in section \ref{sec:direcmet} is a technically complicated endeavor, whereas using the conserved currents is quite easy.

     \section*{Acknowledgements}
     J.H. wants to thank the Finnish Cultural Foundation for providing a part of the funding used for this research.
\nocite{*}

\bibliography{apssamp}% Produces the bibliography via BibTeX.

\end{document}